\documentstyle[11pt]{article}

\begin{document}
\title{\bf Generalized model for human dynamics}
\author{R.~V.~R.~Pandya \thanks{Email: {\sl rvrpturb@uprm.edu}} \\
Department of Mechanical Engineering \\University of Puerto Rico at
Mayaguez\\ Puerto Rico, PR 00681, USA}

\date{\today}
\maketitle

\begin{abstract}
Human dynamics model consistent with our natural ability to perform different activities is put forward by first arguing limitations of the model suggested by Barab\'{a}si (Nature, 435, 207-211, 2005).
\end{abstract}

Humans are capable of performing different types of activities which fill the time available between daily routine basic activities of cleaning themselves, eating and sleeping. The range of activities is substantial but for discussion I mention a few types of activities such as electronic communication, making telephone calls, meeting people and browsing internet etc \cite{barabasi05}. In each type of activity, many events (tasks) are performed at different times, e.g. e-mail reply to a received e-mail is considered as an individual event and many such events belong to electronic communication activity.  Recent observations indicate that the distributions, of waiting time (time taken to reply to a received message) and inter-event time between two consecutive emails during occurrences of many events, depart from the Poisson distribution and have burst characteristics \cite{barabasi05}. This fact led Barab\'{a}si \cite{barabasi05} to suggest a model different than the available models based on Poisson processes and the model indeed captured the observed phenomenon in waiting time distribution. Here first I show weaknesses in Barabási model and then put forward a generalized model for proper description of dynamics of human activities.

In Barab\'{a}si model, each individual has a priority list with $L$ events (tasks) having different priorities governed by parameters $x_i (i=1, 2, …, L)$ chosen from distribution $\rho (x)$. At each time step, the event with highest priority is executed and removed from the list and a new event is added to the list with priority chosen from $\rho (x)$. I elaborate now on the weaknesses of this simple model. This model implicitly assumes that all $L$ events belong to a single activity (e.g. e-mail communication), implying human as servitude and single purpose machine not having any priority at all to perform events of activities other than e-mail reply. Also, the model suggests inter-event time as a constant and equal to the time step, thus incapable of producing proper inter-event time distribution for events belonging to the same activity. 

In the generalized model, each individual has a priority list for $M$ different activities, each having different number of $L_m (m=1, 2, …, M)$ events and $\sum_m L_m=L$ where $L$ is integer constant. The execution times for each event belonging to different $M$ activities are represented by $t_m (m=1, 2, …, M)$. At initial time $(t=0)$ and with some fixed value of $L$, $L_m$ are chosen in consistency with distribution $\omega (y)$ for parameters $y_m$ selected to govern different priorities of $M$ activities. Within each activity $m$, priorities of $L_m$ events belonging to $m$th activity are governed by parameters $x_i  (i=1, 2, …, L_m)$  chosen from distribution $\rho_m(x)$. At each time step, one of the $M$ activities is selected according to the distribution $\omega (y)$ and highest priority event from the selected activity is executed and removed from the list. Subsequently at the end of the execution time, a new event is added to the same activity (say $j$) with priority chosen from corresponding distribution $\rho_j(x)$. The next time step would then be $t_j$ time apart.

This generalized multi-activity model allows presence of events from other activities between two consecutive events of same activity which I consider as possible explanation for observed distribution in inter-event time belonging to the same activity. Though the model implicitly assumes that new events from different activities are available at every time steps, implementing activation of new activities and deactivation of one or a few of existing activities at certain stages of time can be considered as further refinements to the model. Least to mention, events from daily routine basic activities can be added with certainty at appropriate time during the time evolution of human dynamics. Lastly, a simple variant of the generalized model can also be considered in which at each time step highest priority event is selected for execution from all of the events $L$, irrespective of the distribution of priority of different activities. In this case there is a possibility of conflict when more than one event could have identical priority. The priorities of activities can then be invoked to resolve the conflict.


\begin{thebibliography}{1}

\bibitem{barabasi05}
Barabási, A,-L. {\it Nature} {\bf 435}, 207-211 (2005).  

\end{thebibliography}
\end{document}